\pdfoutput=1
\documentclass[sigconf,natbib=true]{acmart}

\usepackage{graphicx}
\usepackage{booktabs}
\usepackage{microtype}
\usepackage{dsfont}
\usepackage{amsmath}
\usepackage{xspace}

\hypersetup{hidelinks}
\AtBeginDocument{%
  }

\setcopyright{none}
\renewcommand\footnotetextcopyrightpermission[1]{}
\copyrightyear{2026}
\acmConference[CIKM '26]{The 35th ACM International Conference on Information and Knowledge Management}{November 7--11, 2026}{Rome, Italy}
\acmYear{2026}
\acmMonth{11}

\newcommand{\vndcg}{\ensuremath{\mathrm{vNDCG}}\xspace}
\newcommand{\bv}{\ensuremath{\mathrm{BV}}\xspace}
\newcommand{\vndcgk}[1]{\ensuremath{\mathrm{vNDCG}@#1}\xspace}
\newcommand{\bvk}[1]{\ensuremath{\mathrm{BV}@#1}\xspace}

\newcommand{\bvarm}[1]{\ensuremath{\mathrm{BV}_{\mathrm{#1}}}}

\begin{document}

\title{Revisiting Generative vs. Dense Retrieval: \\
A Static Value Prior Reverses the Gap}

\author{Renbo Tu}
\authornote{This work was completed during an internship at Shopify.}
\affiliation{\institution{Shopify}}
\affiliation{\institution{University of Toronto}}
\affiliation{\institution{Vector Institute}}

\author{Yang Liu}
\affiliation{\institution{Shopify}}

\author{Shuying Sun}
\authornote{Corresponding author: \href{mailto:shuying.sun@shopify.com}{shuying.sun@shopify.com}.}
\affiliation{\institution{Shopify}}
\expandafter\def\csname typeset@author1\endcsname#1{#1\textsuperscript{1,2,3}}
\expandafter\def\csname typeset@author2\endcsname#1{#1\textsuperscript{1}}
\expandafter\def\csname typeset@author3\endcsname#1{#1\textsuperscript{1}}
\makeatletter
\def\@mkauthors@iii{%
  \author@bx@wd=\textwidth
  \advance\author@bx@wd by -\author@bx@sep
  \divide\author@bx@wd by 3
  \advance\author@bx@wd by -\author@bx@sep
  \gdef\@currentauthors{}%
  \gdef\@currentaffiliation{}%
  \def\@author##1{%
    \ifx\@currentauthors\@empty
      \gdef\@currentauthors{\par##1}%
    \else
      \g@addto@macro\@currentauthors{\par##1}%
    \fi
    \gdef\and{}}%
  \def\email##1##2{}%
  \def\affiliation##1##2{\global\let\and\@typeset@author@bx}%
  \global\setbox\mktitle@bx=\vbox{%
    \noindent\unvbox\mktitle@bx\par\medskip
    \leavevmode\centering
    \addresses\let\and\@typeset@author@bx\and\par
    \smallskip
    {\@affiliationfont
      \textsuperscript{1}\,Shopify\qquad
      \textsuperscript{2}\,University of Toronto\qquad
      \textsuperscript{3}\,Vector Institute\par
      }%
    \bigskip}}%
\makeatother
\renewcommand{\shortauthors}{Tu et al.}

\acmArticleType{Research}
\acmCodeLink{https://anonymous.4open.science/r/gr-dr-release-BCE4}

\begin{abstract}
Generative retrieval (GR) has emerged as a strong candidate generator for sequential recommendation as language models and semantic item identifiers have improved. However, GR is not a uniformly stronger replacement for dense retrieval (DR) in online retail candidate generation. We examine their differences on two Amazon Toys datasets under different metrics. GR leads on the value of the best relevant item among users that both methods cover, while DR leads on exact-target recall and relevant value across the ranked list. Through further analysis, we find that popularity contributes the most to GR's measured value lead. Adding the same static item value to DR in one reranking step reverses this lead and keeps recall above GR. Finally, on an industrial-scale internal e-commerce dataset with LLM-judged relevance, we show that a 14B SID-LLM still underperforms a $47\times$ smaller dense encoder. Our code and artifacts are open-sourced at \url{https://anonymous.4open.science/r/gr-dr-release-BCE4}.

\end{abstract}


\maketitle

\section{Introduction}

\begin{figure}[t]
  \centering
  \includegraphics[width=\columnwidth]{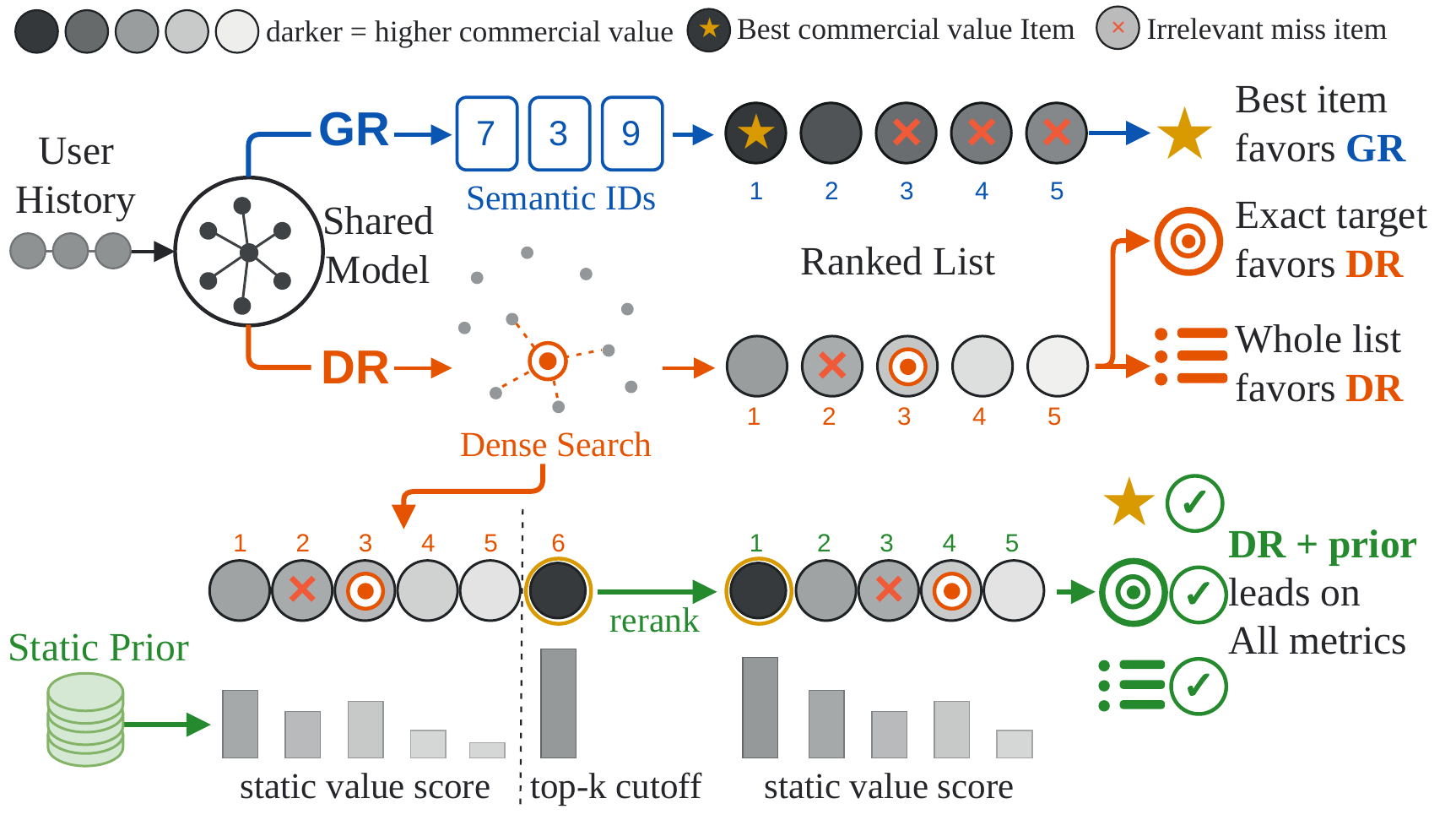}
  \vspace{-15pt}
  \caption{GR and DR use the same user history but generate candidates through semantic-ID decoding and dense search, respectively. GR leads on best-item value, while DR leads on exact-target retrieval and full-list value. A static prior reranks DR candidates and moves a potentially valuable item into the top $k$.}
  \label{fig:overview}
  \vspace{-15pt}
\end{figure}

Dense retrieval (DR) has long been a standard candidate-generation approach for recommendation. It encodes a user's interaction history and catalog items in a shared embedding space, then retrieves nearby items \citep{covington2016youtube,yi2019samplingbias}. Advances in language models and semantic IDs have made generative retrieval (GR) an increasingly viable alternative. GR represents items with semantic IDs and generates the next item's ID one token at a time \citep{decao2021genre,tay2022dsi,wang2022nci,lee2022rqvae,rajput2023tiger}. Recent work has demonstrated GR at industrial scale \citep{zheng2024lcrec,singh2024semanticids,zhou2025onerec,zhou2025onerecv2}, but its use does not establish that it consistently outperforms or can replace DR in candidate generation. As shown in Figure~\ref{fig:overview}, DR typically retrieves the held-out next item more often, while GR's best relevant item has higher value on average for users that both methods cover.

The apparent winner depends on how retrieval is evaluated. This metric choice is particularly crucial in e-commerce candidate generation, where retrieved products differ not only in relevance to the user but also in their commercial value. To understand why, we need to know whether GR's value lead comes from the user's history or from generally valuable items. Recall and target NDCG show whether the held-out item is retrieved and where it ranks. They do not account for other relevant items, their ranks, or their value.

To understand this disagreement, we measure both the value of the best relevant item and the relevant value across the full list. We apply these metrics in controlled GR and DR comparisons on Amazon Toys 2014 and 2023 \citep{mcauley2015image,he2016upsdowns,hou2024blair,rajput2023tiger,yang2024liger}. The metric disagreement persists on both public e-commerce datasets. Our analysis shows that popularity is the largest component of GR's measured value lead. Among users for whom GR wins, popularity contributes 49.8\% of its value lead on 2014 and 65.4\% on 2023. Reranking DR with the static item value score reverses the best-item value gap while preserving a recall lead over GR. Further checks with cold targets, substitute credit, a random control, and graded industrial relevance do not reveal a consistent GR advantage. Our main contributions are:

\begin{itemize}
  \item We define complementary metrics for sequential recommendation in e-commerce settings and show that they favor different methods in the GR--DR comparison.
  \item We decompose GR's measured value lead and find that popularity is its largest component on both public datasets.
  \item We propose a training-free value prior, applied to DR as a single rerank of its retrieval results. This reverses GR's lead in the value of the best relevant item while preserving DR's recall lead.

\end{itemize}

\section{Related Work}

\paragraph{The Generative Turn for Retrieval Systems}
Prior to decoder-only approaches, sequential recommendation models for retrieval were prevalent in production \citep{covington2016youtube, yi2019samplingbias}. The canonical dense retrieval (DR) is therefore defined by bidirectional sequential models for next-item prediction \citep{hidasi2016gru4rec, kang2018sasrec, sun2019bert4rec}. Generative retrieval (GR) enters the foreground as a constrained autoregressive decoding mechanism to generate identifier tokens, semantic IDs (SIDs), mappable to an underlying item, whose codebook construction follows RQ-VAE \citep{decao2021genre, tay2022dsi, wang2022nci, lee2022rqvae}. SID generative retrieval benefits from the rise of LLMs and demonstrates generalization, establishing itself in industry applications \citep{rajput2023tiger, zheng2024lcrec, singh2024semanticids, zhou2025onerec, zhou2025onerecv2}. Note that GR also encompasses non-SID approaches such as P5 and HSTU \citep{geng2022p5, zhai2024hstu, liu2026trope}. Our controlled DR--GR comparison is grounded in LIGER, while research on GR's scaling difficulties and DR's capacity limits bounds our conclusions without determining them \citep{yang2024liger, pradeep2023scaling, weller2026limit}.

\paragraph{Difficulty of Retrieval Evaluation}
Offline recommendation evaluation suffers from sensitivity to missing feedback, candidate sampling, and temporal splitting decisions \citep{steck2010mnar, cremonesi2010topn, krichene2020sampled, canamares2020target, dallmann2021sampling, ji2023leakage}. Audits of these decisions reveal that they can change recommendation rankings, while recent studies identify GR artifacts from SID collisions and benchmark shortcuts \citep{ferraridacrema2019progress, hidasi2023flaws, sun2023freshlook, zhang2026sidreliability, han2026shortcuts}. Our contribution identifies substitute non-identifiability under exact-target metrics and shows how it worsens with catalog scale.

\paragraph{Popularity Bias Is a Static Prior} In recommendation, the popularity of certain products is deeply associated with benchmark accuracy \citep{steck2011popularity, jannach2015biases}. However, popularity also presents itself as a bias that degrades exposure and fairness for non-popular items \citep{abdollahpouri2019unfairness, klimashevskaia2024survey}. Popularity bias in GR becomes prominent due to inherent bias within LLMs \citep{zhang2023chatgptfair, yin2026echoes, fan2026crab}. While these works mitigate popularity bias, our results show that, under a same-model GR--DR comparison, the advantage from popularity is replaceable by a user-independent prior that can be transferred to DR without decoding.

\section{Controlled Comparison Setup}
\label{sec:setup}

\paragraph{Problem Definition}
In the e-commerce setting, for each user $u$, let $h_u=(i_1,\ldots,i_t)$ denote the observed user--product interaction history and $y_u=i_{t+1}$ the held-out next item. A candidate generator $x$ uses $h_u$ to produce the ranked list
\[
L_x^k(u)=\bigl(i_{x,u}^{(1)},\ldots,i_{x,u}^{(k)}\bigr),
\]
where $i_{x,u}^{(j)}$ is the item at rank $j$.

\paragraph{Models}
We compare GR and DR using the TIGER/LIGER framework \citep{rajput2023tiger,yang2024liger}. Both arms use the same interaction data, item catalog, and jointly trained T5 encoder--decoder. GR decodes valid semantic IDs through prefix-constrained beam search, while DR ranks items by cosine similarity using the encoder representation of the same history. Thus, the arms differ mainly in their retrieval readout.

\paragraph{Datasets}

Our public evaluation uses the 2014 and 2023 releases of Amazon Toys. The 2014 dataset contains 11,924 items and 19,412 user histories under a leave-last-out split. Among the held-out targets, 19,116 were seen during training and 296 were absent from training. Our results reproduce the LIGER Recall@10 values, while our GR recall exceeds the published TIGER result. The 2023 catalog is $13.6\times$ larger, which lets us test whether the observed pattern and the proposed intervention survive a substantial increase in catalog scale \citep{hou2024blair}. We transfer the value weights and reranking coefficient from 2014 to 2023 without change unless stated otherwise.

\section{Evaluation Metrics}
\label{sec:evaluation}

We compare GR and DR by asking \textit{which retrieved items are relevant} and \textit{how much  value they carry}. A relevance labeling rule $q_u(i)\in[0,1]$ assigns evaluation credit after ranking and is not a model score. It may credit only the held-out item, include substitute candidates, or assign partial credit through graded judgments. In contrast, $v_i\in[0,1]$ measures a static property of item $i$. For any rule $q$, the relevant set with $k$ candidates is defined as
\begin{equation*}
R_x^k(u;q)=\{i\in L_x^k(u)\mid q_u(i)>0\}.
\end{equation*}
This definition keeps relevance and item value separate across evaluation protocols. In this section, we first define exact-target and category-based relevance, then introduce static catalog value, and finally combine them at the best-item and whole-list levels.

\subsection{Exact-Target Retrieval}
The conventional next-item protocol assigns credit only to the held-out item, as defined by $q_u^{\mathrm{exact}}(i)=\mathds{1}[i=y_u]$. Recall@$k$ measures whether this item appears in the top-$k$ results.
\begin{equation*}
\mathrm{Recall}_x@k
=\frac{1}{|\mathcal U|}\sum_{u\in\mathcal U}
\mathds{1}[y_u\in L_x^k(u)].
\end{equation*}
NDCG@$k$ evaluates the same target with a rank discount \citep{jarvelin2002ndcg}. Both metrics measure recovery of the observed next interaction and give no credit to other potentially useful items.

\subsection{Beyond Exact Match: Substitute Blindness}
\label{subsec:substitute-definition}
Exact-target evaluation cannot distinguish a substitute from an irrelevant item because both receive zero credit. We refer to this limitation as \emph{substitute blindness}. Therefore, we define a coarse proxy for substitute-aware relevance using the shared leaf category as
\begin{equation*}
q_u^{\mathrm{cat}}(i)
=\mathds{1}[\mathrm{leaf}(i)=\mathrm{leaf}(y_u)],
\quad
S(u)=\{i\neq y_u:q_u^{\mathrm{cat}}(i)=1\},
\end{equation*}
where $\mathrm{leaf}(i)$ and $\mathrm{leaf}(y_u)$ identify the leaf categories of candidate item $i$ and the held-out target. An item receives $q_u^{\mathrm{cat}}(i)=1$ when the two categories match. A larger substitute set $S(u)$ makes an exact miss less informative, but category overlap alone does not establish that two items are substitutes. We therefore stress-test this proxy using co-purchase credit, a mass-matched random-credit control, and LLM-judged graded relevance in Section~\ref{sec:robustness}.

\subsection{Static Item Value}
\label{subsec:item-value}

Relevance determines whether an item is useful for a given user. In e-commerce, products also differ in static catalog attributes that may matter to the platform. We summarize these attributes with a transparent item-value proxy
\begin{equation}
v_i(\mathbf w)
=w_p\,\mathrm{pop}_i+w_r\,\mathrm{rate}_i+w_b\,\mathrm{brand}_i,
\quad w_p+w_r+w_b=1.
\label{eq:value}
\end{equation}
Here, $\mathrm{pop}_i$, $\mathrm{rate}_i$, and $\mathrm{brand}_i$ are the catalog-percentile ranks of log popularity, the Wilson lower bound on star rating, and brand footprint, respectively \citep{wilson1927score}. All weights are nonnegative. Unless otherwise stated, we use the default weights $\mathbf w=(0.25,0.25,0.50)$. This score is a transparent proxy for item value, not a universal measure of recommendation utility. In the sensitivity analysis, we vary $\mathbf w$ over nonnegative mixtures whose components sum to one.
Unlike $q_u(i)$, which varies across users and relevance labeling rules, $v_i(\mathbf w)$ is computed once for each catalog item.

\subsection{Best-Item Value vs. Whole-List Value}

Having defined relevance $q$ and item value $v$ separately, we evaluate each ranked list using two complementary metrics.

\paragraph{Conditional Best-Item Value (BV)}
BV asks how valuable the best relevant item returned by method $x$ is for user $u$.
\begin{equation*}
\bvarm{x}^k(u;q,\mathbf w)
=\max_{i\in R_x^k(u;q)}v_i(\mathbf w).
\end{equation*}
In our main analysis, BV uses the binary category rule $q^{\mathrm{cat}}$ from Section~\ref{subsec:substitute-definition} to select relevant items but does not weight them by relevance magnitude. Let $\mathcal C_{x,a}^k$ denote the users for whom both method $x$ and comparator $a$ retrieve at least one relevant item. We report their average gap as
\begin{equation*}
\Delta\mathrm{BV}_{x-a}@k
=\frac{1}{|\mathcal C_{x,a}^k|}
\sum_{u\in\mathcal C_{x,a}^k}
\left[\bvarm{x}^k(u)-\bvarm{a}^k(u)\right],
\end{equation*}
where $q$ and $\mathbf w$ are omitted for readability. We compare the BV scores of methods $x$ and $a$ for each user $u$. A difference above $\tau=0.05$ is a win for $x$, a difference below $-\tau$ is a win for $a$, and all other differences are ties.

\paragraph{Whole-List Value (\vndcg)}
BV focuses on the best relevant item. Its comparison includes only users with a relevant item from both methods. To score the full top-$k$ list for every user, we use NDCG with item value weights (\vndcg). Each item contributes its relevance times its value, with less credit at lower ranks. For method $x$ and user $u$, let $i_j=i_{x,u}^{(j)}$ be the item at rank $j$. The list score is
\begin{equation*}
\mathrm{vDCG}_x^k(u)
=\sum_{j=1}^{k}
\frac{q_u(i_j)v_{i_j}(\mathbf w)}{\log_2(j+1)}.
\end{equation*}
For each user, we pool the items returned by all methods and rank them by relevance times value, $q_u(i)v_i(\mathbf w)$. The discounted score of the top $k$ items is the ideal score $\mathrm{IDCG}^k(u)$. We divide each method's score by this shared score, then average across users:
\begin{equation*}
\mathrm{vNDCG}_x@k
=\frac{1}{|\mathcal U|}\sum_{u\in\mathcal U}
\frac{\mathrm{vDCG}_x^k(u)}{\mathrm{IDCG}^k(u)}.
\end{equation*}

\begin{table}[t]
\centering
\vspace{-4pt}
\caption{Amazon Toys results. Positive $\Delta$BV@50 values favor the listed method over GR. Best values are in bold.}
\vspace{-8pt}
\label{tab:main-results}
\footnotesize
\setlength{\tabcolsep}{3.5pt}
\resizebox{\columnwidth}{!}{%
\begin{tabular}{@{}lrrrrrrrr@{}}
\toprule
 & \multicolumn{2}{c}{Recall@50}
 & \multicolumn{2}{c}{NDCG@50}
 & \multicolumn{2}{c}{$\Delta$BV@50}
 & \multicolumn{2}{c}{\vndcgk{50}} \\
\cmidrule(lr){2-3}\cmidrule(lr){4-5}\cmidrule(lr){6-7}\cmidrule(l){8-9}
Method & 2014 & 2023 & 2014 & 2023 & 2014 & 2023 & 2014 & 2023 \\
\midrule
GR         & 0.167          & 0.052          & 0.056          & 0.017          & ---               & ---               & 0.215          & 0.160 \\
DR         & \textbf{0.228} & \textbf{0.084} & \textbf{0.077} & \textbf{0.029} & $-0.020$          & $-0.016$          & 0.222          & 0.168 \\
DR+prior   & 0.210          & 0.083          & 0.072          & 0.028          & $\mathbf{+0.023}$ & $\mathbf{+0.006}$ & \textbf{0.240} & \textbf{0.175} \\
\bottomrule
\end{tabular}}
\vspace{-8pt}
\end{table}

If the ideal score is zero for a user, their \vndcg score is zero. All methods use the same ideal score for each user, so their scores can be compared within the fixed pool. Absolute scores from different method pools should not be compared directly. To check whether coverage alone explains the metric gap, we also compare BV over all users, assigning a BV of zero when a method retrieves no relevant item.

\section{From Metric Dependency to a Static Prior}
\label{sec:main-analysis}

We now apply these complementary metrics in a three-step analysis.
We first show that GR leads on BV while DR leads on \vndcg. We then measure how popularity, rating, and brand contribute to GR's value advantage. Finally, we rerank DR with the same static item value and test whether the gap reverses.

\subsection{The Metric-Dependent GR--DR Gap}
\label{sec:metric-gap}

\begin{figure}[t]
  \centering
  \vspace{-6pt}
  \includegraphics[width=\columnwidth]{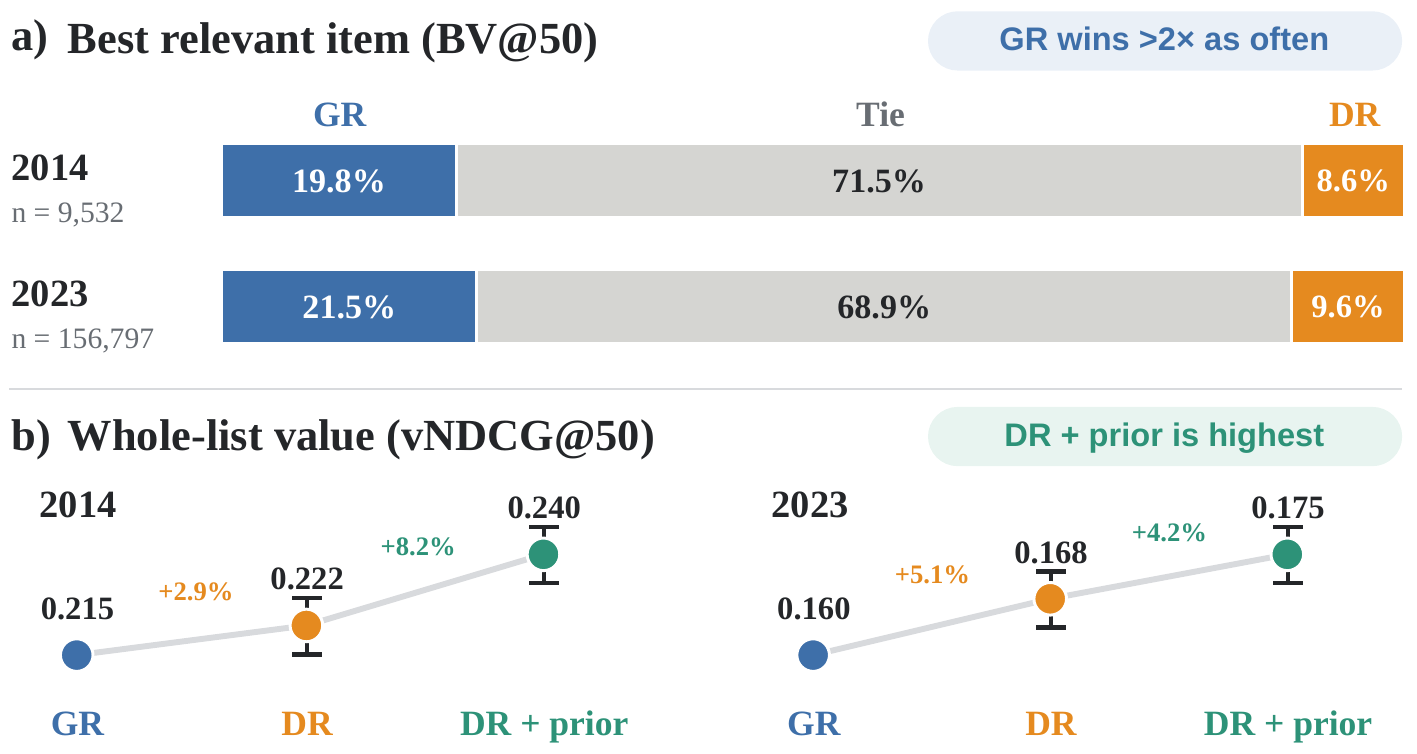}
  \vspace{-18pt}
  \caption{Metric choice reverses the result. (a) GR wins more often under \bvk{50} on both datasets. (b) Under \vndcgk{50}, DR scores higher than GR, and DR+prior scores highest. Error bars show standard errors for adjacent differences.}
  \label{fig:headline}
  \vspace{-14pt}
\end{figure}

\begin{figure}[t]
  \centering
  \vspace{-10pt}
  \includegraphics[width=\columnwidth]{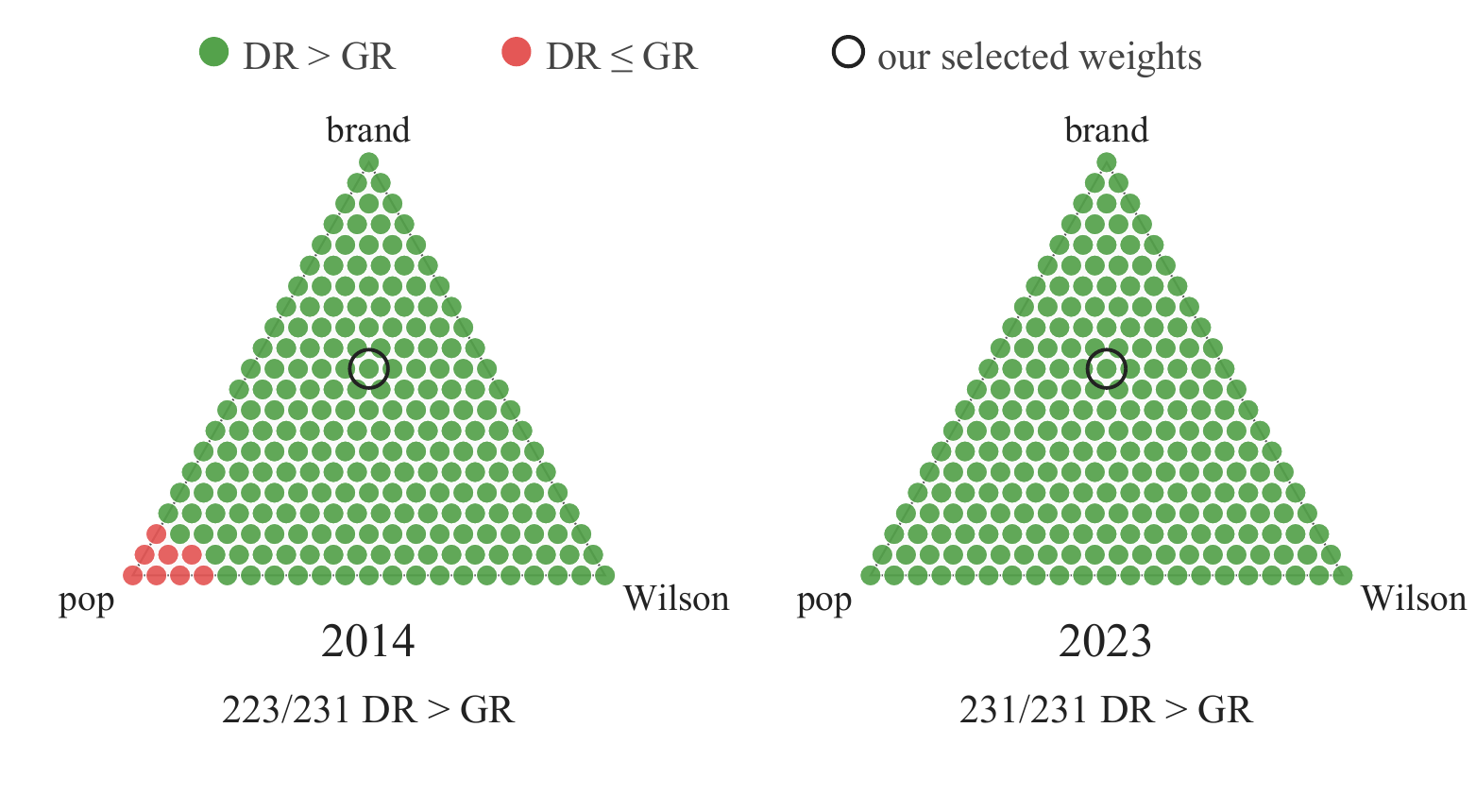}
  \vspace{-34pt}
  \caption{Sign of the raw DR-vs-GR \vndcgk{50} gap over the value-weight simplex on the two Amazon Toys datasets. The grid step is 0.05.}
  \label{fig:simplex}
  \vspace{-14pt}
\end{figure}

We begin with the empirical disagreement.
GR leads on BV, while raw DR leads on \vndcgk{50}, Recall@50, and target NDCG@50 on both datasets, as shown in Table~\ref{tab:main-results}. GR wins the paired BV comparison more than twice as often as raw DR (Figure~\ref{fig:headline}(a)), and its lead remains when users without a relevant retrieved item receive zero. DR's \vndcgk{50} lead is 2.9\% on 2014 and 5.1\% on 2023, with confidence intervals above zero (Figure~\ref{fig:headline}(b)). Value weighting compresses DR's relevance advantage by an order of magnitude. We conduct a full sweep over value weights to demonstrate robustness. Across the 231-point weight simplex shown in Figure~\ref{fig:simplex}, the sign of the \vndcg gap holds on 223 points on 2014, failing only in popularity-dominated edge cases, and on all 231 points on 2023.

\subsection{Diagnosing GR's Value Advantage}
\label{sec:diagnosis}

Among users for whom GR wins on \bv, we use Equation~\eqref{eq:value} to break its value lead over DR into popularity, rating, and brand. Popularity contributes 49.8\% of the lead on 2014 and 65.4\% on 2023. It is the largest component on both datasets, and its share is higher on the larger 2023 catalog. Rating and brand contribute the remainder, but neither provides the same cross-dataset dominance.

This pattern narrows the interpretation of GR's value edge. Each value component is computed once per item and does not depend on the current history. The measured lift is therefore largely captured by static item-level properties, especially popularity. Repeated exposure in training offers a possible explanation for GR's value lead. The released 2014 popularity score correlates with how often held-out targets appear in training histories (Spearman \(\rho=0.582\)). At least ten training histories contain 50.3\% of held-out targets on 2014 and 75.1\% on 2023. GR may learn a preference for frequently seen items while predicting their semantic IDs. This could contribute to its lead because popularity is part of the item value score. This explanation yields a falsifiable implication. If static item properties account for the operational advantage, then explicitly adding the same information to DR should reproduce it without autoregressive decoding.

\subsection{Transferring the Static Prior to DR}
\label{sec:prior}

To test whether static item value can close GR's lead in \bv, we rerank DR's top 200 candidates for each user. We call this method DR+prior. DR+prior scales the DR similarity scores to $[0,1]$ within each pool, then averages each score with the item's value $v_i(\mathbf w)$ using the same default weights as \bv (Section~\ref{subsec:item-value}). Because \bv uses the same item value, this test checks whether an explicit catalog prior can reproduce GR's measured lead. This requires no retraining or decoder at serving time. We score each candidate $i$ in user $u$'s pool $P_u$ by
\begin{equation*}
\sigma_u(i) \;=\; (1-\alpha)\,\tilde s_u(i) \;+\; \alpha\, v_i,
\qquad \alpha=0.5, \quad i \in P_u,
\end{equation*}
where $\tilde s_u(i)$ is the normalized DR score.

As a result, adding the prior turns DR's \bvk{50} deficit against GR into a lead. On 2014, DR moves from trailing by $0.020$ to leading by $0.023$. On 2023, it moves from trailing by $0.016$ to leading by $0.006$. As shown in Table~\ref{tab:main-results}, DR+prior also has the highest \vndcgk{50} on both datasets, at 0.240 and 0.175. Its Recall@50 declines relative to raw DR but stays above GR.

\section{Robustness Beyond Exact-Match Evaluation}
\label{sec:robustness}

The public results show that the item prior closes GR's value gap. We now test whether GR does better on cold targets or when useful alternatives to the held-out target receive credit. For the latter, we use public substitute evidence and graded judgments on an industrial recommendation dataset.

\subsection{Cold Items}

We test whether generation helps retrieve targets absent from the training interactions. GR finds none of the 296 cold targets on 2014 and 5 of 1,899 on 2023. DR finds 47 and 27, while DR+prior finds 36 and 21, respectively. Both DR methods find more cold targets than GR on each dataset.

\subsection{Substitute Credit and Graded Relevance}

An exact-target miss does not tell us whether a list contains a useful alternative. For the median user, the held-out target has 157 same-leaf alternatives on 2014 and 1,046 on 2023, a $6.7\times$ increase. Among DR's 8,635 exact-target misses on 2014, 58.2\% include an item from that category in the top 50 while the target ranks beyond 200. In another 25.7\%, the target ranks between 50 and 200. GR finds the exact target in only 748 of these DR misses, so the misses alone do not show an advantage for GR.

We give binary credit to items linked to the target in Amazon co-purchase data (Table~\ref{tab:substitutes}). In this test, a 0.0447 lead for DR under exact Recall@50 becomes a 0.0183 lead for GR. For the binary test, randomly chosen items given the same amount of credit move the gap at least as far toward GR. With graded NDCG, DR leads again by 0.0153. The binary reversal therefore does not establish that GR retrieves better alternatives.
Leaf-category overlap and co-purchase links are imperfect evidence of usefulness. We next test graded relevance on the industrial dataset.

\begin{table}[t]
\centering
\caption{Substitute credit on 2014. The last column shows the extra shift toward GR from co-purchase credit compared with random credit of equal total weight.}
\vspace{-8pt}
\label{tab:substitutes}
\begin{tabular}{@{}lcc@{}}
\toprule
 & DR--GR & Real--random credit \\
\midrule
Recall@50, exact only   & $+0.0447$ & --- \\
$+$ co-purchase credit & $-0.0183$ & $-0.0190$ \\
$+$ graded NDCG@50     & $+0.0153$ & $+0.0111$ \\
\bottomrule
\end{tabular}
\vspace{-8pt}
\end{table}

\subsection{Industrial LLM-Judged Relevance}

The industrial evaluation uses real e-commerce interaction histories and a product catalog. GR uses a Qwen3-14B model that generates semantic IDs, while DR uses a GTE encoder with about 300 million parameters. GR has roughly $47\times$ as many parameters. A substitute-aware LLM judge calibrated with 250 human labels scores each candidate's relevance from zero to one, allowing partial credit for useful alternatives. Because the models differ in backbone and size, this evaluation tests whether the public result also holds in this setting. It cannot isolate the effect of retrieval architecture. Table~\ref{tab:internal} shows that giving useful alternatives graded credit preserves the order GR $<$ DR $<$ DR+prior on the industrial dataset. The prior raises DR's NDCG@50 by 11.8\%. %

\begin{table}[t]
\centering
\caption{Industrial dataset evaluation under an LLM judge.}
\vspace{-8pt}
\label{tab:internal}
\begin{tabular}{@{}lrrr@{}}
\toprule
Population (NDCG@50) & GR & DR & DR+prior \\
\midrule
Full evaluation set ($n=5{,}998$) & 0.453 & 0.474 & \textbf{0.530} \\
\bottomrule
\end{tabular}
\vspace{-8pt}
\end{table}

\section{Conclusion}

In this paper, we revisited GR and DR in a controlled e-commerce comparison and found two distinct parts to GR's measured advantage. First, metric choice changes the apparent winner. GR leads on the value of the best relevant item, while DR leads on exact-target recall and relevant value across the ranked list on both public datasets. Second, popularity makes the largest contribution to GR's value lead. Adding the same static item value to DR in one reranking step reverses this gap while preserving DR's recall lead. Cold-target and substitute-credit checks do not reveal a broader GR advantage. Graded judgments on an industrial dataset place DR above a larger GR model and DR with the prior above both. These results make DR with a static value prior a useful baseline for future GR systems.

\newpage
\bibliographystyle{ACM-Reference-Format}
\bibliography{references}

\end{document}